\documentclass[
aps, prapplied, reprint, superscriptaddress, amsmath, amssymb, 10pt, longbibliography, nobibnotes
]{revtex4-2}

\usepackage{graphicx}
\usepackage{dcolumn}
\usepackage{bm}
\usepackage{textcomp}
\usepackage{gensymb}
\usepackage{siunitx}
\usepackage{isotope}
\usepackage[T1,T2A]{fontenc}
\usepackage[utf8]{inputenc}
\usepackage{xcolor}

\DeclareUnicodeCharacter{2212}{{–}}
\DeclareUnicodeCharacter{03BB}{{$\lambda$}}
\DeclareUnicodeCharacter{2009}{{ }}
\DeclareUnicodeCharacter{2033}{{"}}
\DeclareUnicodeCharacter{02B9}{{'}}

\begin{document}

\title{\textbf{Active compensation of the AC Stark shift in a two-photon rubidium optical frequency reference using power modulation}}

\author{Yorick Andeweg}
\email{Contact author: yorick.andeweg@colorado.edu.}
\affiliation{National Institute of Standards and Technology, Boulder, Colorado, United States}
\affiliation{University of Colorado Boulder, Boulder, Colorado, United States}

\author{John Kitching}
\affiliation{National Institute of Standards and Technology, Boulder, Colorado, United States}
\author{Matthew T. Hummon}
\affiliation{National Institute of Standards and Technology, Boulder, Colorado, United States}

\begin{abstract}
We implement a feedback protocol to suppress the AC Stark shift in a two-photon rubidium optical frequency reference, reducing its sensitivity to optical power variations by a factor of 1000. This method alleviates the tradeoff between short-term and long-term stability imposed by the AC Stark shift, enabling us to simultaneously achieve instabilities of $3\times10^{-14}$ at 1 s and $2\times10^{-14}$ at $10^4$ s. We also quantitatively describe, and experimentally explore, a stability limit imposed on clocks using this method by frequency noise on the local oscillator.
\end{abstract}

\maketitle

\section{Introduction}
Optical atomic clocks represent the state of the art in precision timekeeping, with several advanced, laboratory-based optical clocks achieving a systematic fractional frequency instability at the $10^{-18}$ level \cite{brewer_27mathrm_2019, aeppli_clock_2024, mcgrew_atomic_2018, ohmae_transportable_2021}. Currently, almost all atomic clocks deployed in the field for navigation, communication, and sensing applications operate at microwave frequencies, but advances in compact frequency combs \cite{lagatsky_high-performance_2023, timmers_ruggedized_2024, zhang_ultrabroadband_2025} promise to usher in a new generation of deployable clocks based on optical transitions. One candidate for such technology is the molecular iodine clock \cite{hackel_observation_1975, ye_molecular_2001}, which has demonstrated performance comparable to that of a hydrogen maser while aboard a naval ship \cite{roslund_optical_2024}. Another candidate is based on the two-photon 5S$_\frac{1}{2}$ $\rightarrow$ 5D$_\frac{5}{2}$ transition at 778 nm in rubidium \cite{nez_optical_1993, Millerioux1994a, Grove1995, danielli_frequency_2000, Edwards2005, Wu2014a, Xia2016, terra_ultra-stable_2016, martin_compact_2018, lemke_measurement_2022, nguyen_field-programmable_2024, erickson_atomic_2024, ahern_tailoring_2025}.

The two-photon rubidium transition shows strong potential for use in a high-performance secondary optical frequency reference (OFR). Its multiphoton nature allows two overlapping, counterpropagating excitation beams to perform Doppler-free spectroscopy in a warm vapor \cite{cagnac_multiphotonic_1973}, requiring less complexity than cold atom systems. Furthermore, the transition benefits from a high quality factor, a clear resonance signal provided by straightforward fluorescence detection at 420 nm, and a fortuitous wavelength that enables the use of mature 1550 nm telecommunications components together with second harmonic generation (SHG) \cite{poulin_absolute_1997}. Finally, thanks to advances in microfabricated vapor cells \cite{liew_microfabricated_2004} and integrated photonics, OFRs and clocks based on this transition can be engineered into packages with small size, weight, and power consumption \cite{burke_compact_2016, newman_architecture_2019, Maurice2020, newman_high-performance_2021, hilton_demonstration_2025}. They therefore show strong potential for portability and robust performance in non-laboratory environments.

However, the two-photon transition in rubidium suffers from a large AC Stark shift from the excitation beams. Two-photon rubidium OFRs are therefore sensitive to fluctuations in both the beams' intensity and their degree of overlap \cite{martin_frequency_2019}, which are difficult to stabilize at long timescales. This introduces an inherent tradeoff between short-term stability, which benefits from intense excitation light, and long-term stability, which suffers from it. For practical choices of operating parameters, the AC Stark shift often limits long-term performance \cite{martin_compact_2018, newman_high-performance_2021, ahern_tailoring_2025}.

Alleviating the tradeoff imposed on this platform by the AC Stark shift is an area of active research. Several methods have been proposed, including a scheme in which the error signal generated by the atomic interrogation is locked to a nonzero value \cite{li_frequency_2024}, a scheme in which two vapor cells are simultaneously interrogated at different intensities \cite{li_dual-interrogation_2024}, and schemes in which two lasers at different wavelengths are employed to cancel each others' AC Stark shifts, with the ratio of their intensities either nominally fixed \cite{blum_light_2024} or actively stabilized by a secondary feedback loop \cite{gerginov_two-photon_2018}. The methods described in Refs. \cite{li_dual-interrogation_2024} and \cite{blum_light_2024} rely on the ability to accurately measure the relative intensities of two different beams, and the method described in Ref. \cite{li_frequency_2024} introduces sensitivity to variations in other parameters. Nevertheless, these methods have been demonstrated to suppress the effects of large, intentionally applied optical power variations. The method described in Ref. \cite{gerginov_two-photon_2018} shows promise but has so far been only been demonstrated as a proof of concept rather than a closed-loop operational clock.

In this work, we implement a protocol called the auto-compensated shift (ACS) \cite{yudin_general_2020} to suppress the AC Stark shift in a two-photon rubidium OFR. This protocol was designed for implementation in a generic continuous-wave (CW) clock and has been previously demonstrated in a microwave clock based on coherent population trapping (CPT) \cite{abdel_hafiz_protocol_2020}. ACS (along with the method described in Ref. \cite{gerginov_two-photon_2018}) belongs to a family of related techniques for compensating clock shifts that also includes Autobalanced Ramsey spectroscopy \cite{sanner_autobalanced_2018, abdel_hafiz_toward_2018, abdel_hafiz_symmetric_2018, shuker_ramsey_2019, abdel_hafiz_light-shift_2022} and a number of techniques for optically pumped microwave clocks \cite{arditi_application_1975, hashimoto_novel_1990, mcguyer_simple_2009, calosso_laser-frequency_2024} and CW CPT clocks \cite{shah_continuous_2006, yudin_method_2021, radnatarov_active_2023}. These techniques have been collectively described as ``two-loop methods'' \cite{yudin_generalized_2018, yudin_general_2020} and are united by their employment of two simultaneous feedback loops, both deriving their error signals from atomic interrogation: a primary loop that stabilizes the frequency of the local oscillator (LO) to the atomic transition frequency, and a secondary loop that stabilizes an additional experimental parameter to compensate some perturbation or clock shift. In previous work, the additional experimental parameter has been called the ``concomitant'' parameter and is often denoted by $\xi$ \cite{yudin_generalized_2018, yudin_general_2020}. The stipulation that both error signals be derived from atomic interrogation is what distinguishes the secondary feedback loop in one of these so-called two-loop methods from any number of other independent feedback loops that may exist in a clock system, such as the stabilization of a component's temperature using a thermistor. The distinction matters because extracting information from the atoms eases requirements on the ability to directly measure and stabilize the offending perturbation. The term ``two-loop method'' could even be extended to refer to methods where the second loop is not implemented in hardware, but rather activated infrequently by hand during brief pauses in a clock's operation, such as the occasional tuning of a magnetic compensation field applied to a lattice clock to minimize its transition linewidth \cite{boyd_sr_2007}.

A consequence of implementing ACS is that the secondary feedback loop provides a new avenue for frequency noise on the LO to couple onto the clock's stable output, imposing a new limit on the clock's stability. Authors investigating other two-loop methods have remarked on this phenomenon \cite{gerginov_two-photon_2018, calosso_laser-frequency_2024, li_dual-interrogation_2024}. In this work, we quantitatively describe the stability limit imposed on ACS clocks by LO frequency noise. We systematically vary the parameters of our ACS implementation to experimentally investigate this stability limit, and the results are reasonably consistent with the model we present.

\begin{figure}
    \centering
    \includegraphics[width=3.4in]{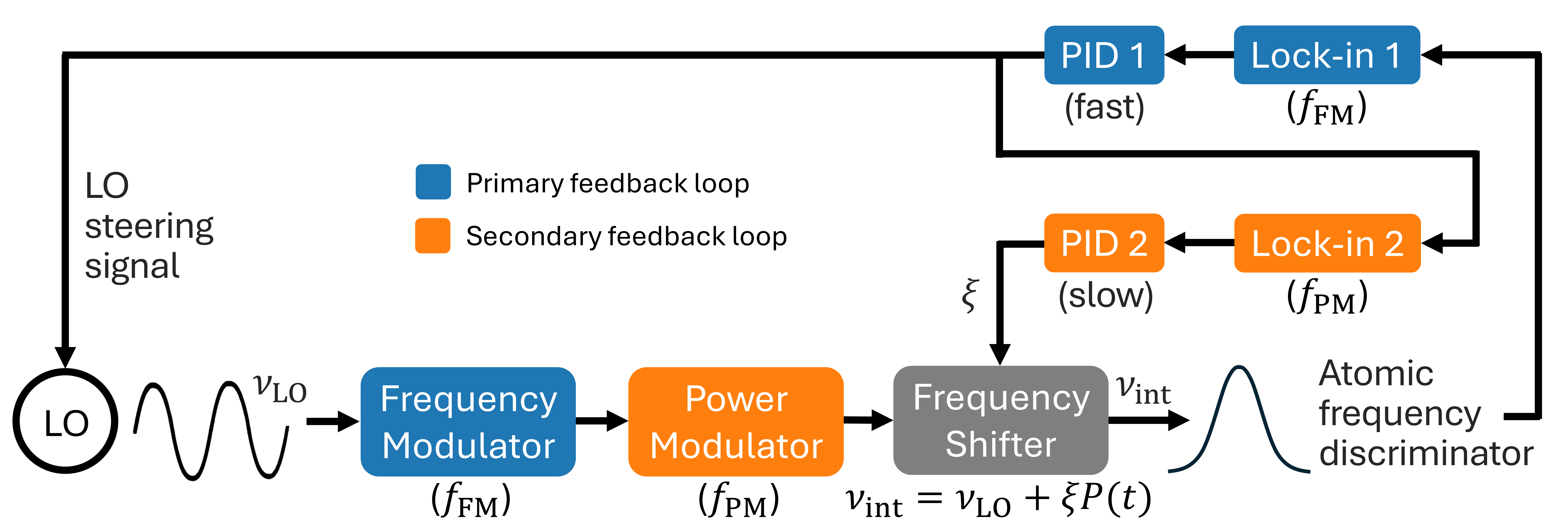}
    \caption{A high-level block diagram of the ACS method for suppressing the AC Stark shift in a generic CW frequency reference. PID = proportional/ integral/ derivative controller.}
    \label{fig_abstracted_diagram}
\end{figure}

\section{Principle of ACS}
ACS is a two-loop method for compensating the AC Stark shift due to the interrogation field in a generic CW frequency reference. Its working principle is described here. The AC Stark shift is assumed to be linear with the interrogation intensity. A block diagram of the ACS method is shown in Fig. \ref{fig_abstracted_diagram}.

The defining characteristic of ACS is the introduction of a frequency shifter to the interrogation beam, creating a controllable offset between the LO frequency $\nu_\text{LO}$ before the shifter and the interrogation frequency $\nu_\text{int}$ after the shifter. This component is programmed to apply a frequency shift proportional to the instantaneous measured interrogation power $P(t)$:
\begin{equation}
    \label{eq_shift}
    \nu_\text{int} = \nu_\text{LO} + \xi P(t).
\end{equation}
where the constant of proportionality $\xi$ is controllable. $\xi$ is nominally real-valued; that is, there is no intentional phase shift between the power modulation and the applied frequency shift. The clock's primary loop operates conventionally, steering $\nu_\text{LO}$ to lock $\nu_\text{int}$ to the (AC Stark-shifted) atomic transition frequency with sufficient bandwidth for robust operation. Meanwhile, the secondary feedback loop controls $\xi$, slowly and continually adjusting its value to match the transition's AC Stark shift coefficient. If $\xi$ is adjusted correctly, then the applied frequency shift absorbs the AC Stark shift, and $\nu_\text{LO}$ becomes insensitive to variations in the interrogation power.

In a conventional frequency reference, the measured interrogation power $P$ is stabilized to a constant value in order to minimize AC Stark shift fluctuations. In ACS, the power is modulated, for example sinusoidally:
\begin{equation}
    \label{eq_power_modulation}
    P(t) = P_0 [1 + A \sin(2 \pi f_\text{PM} t)],
\end{equation}
where $P_0$ is the average interrogation power, $A$ is the modulation amplitude, $t$ is time, and $f_\text{PM}$ is the power modulation frequency, which should be slower than the bandwidth of the primary feedback loop. The primary loop should have sufficient dynamic range for the LO to remain locked through a complete cycle of the power modulation. If $\xi$ is not matched to the AC Stark shift coefficient, the applied modulation causes a tone at $f_\text{PM}$ to appear on the LO steering signal. A lock-in amplifier detects this tone and provides the error signal for the secondary feedback loop, which controls $\xi$ with a response time $T \gg 1/f_\text{PM}$ to make the tone vanish. The lock-in amplifier's phase should be chosen so that when $\xi$ is 0, the measured response to the applied modulation lies as much as possible in one quadrature, which can be achieved by minimizing the signal on the other quadrature.

ACS is tolerant to slowly-varying multiplicative errors in the power measurement that would, in a conventional power stabilization setup, lead to OFR frequency fluctuations. As long as the \textit{ratio} between the maximum power and the minimum power during each modulation cycle is well-measured, the secondary feedback loop ensures that $\nu_\text{LO}$ is extrapolated to the zero-AC Stark shift transition frequency.

\begin{figure}
    \centering
    \includegraphics[width=3.4in]{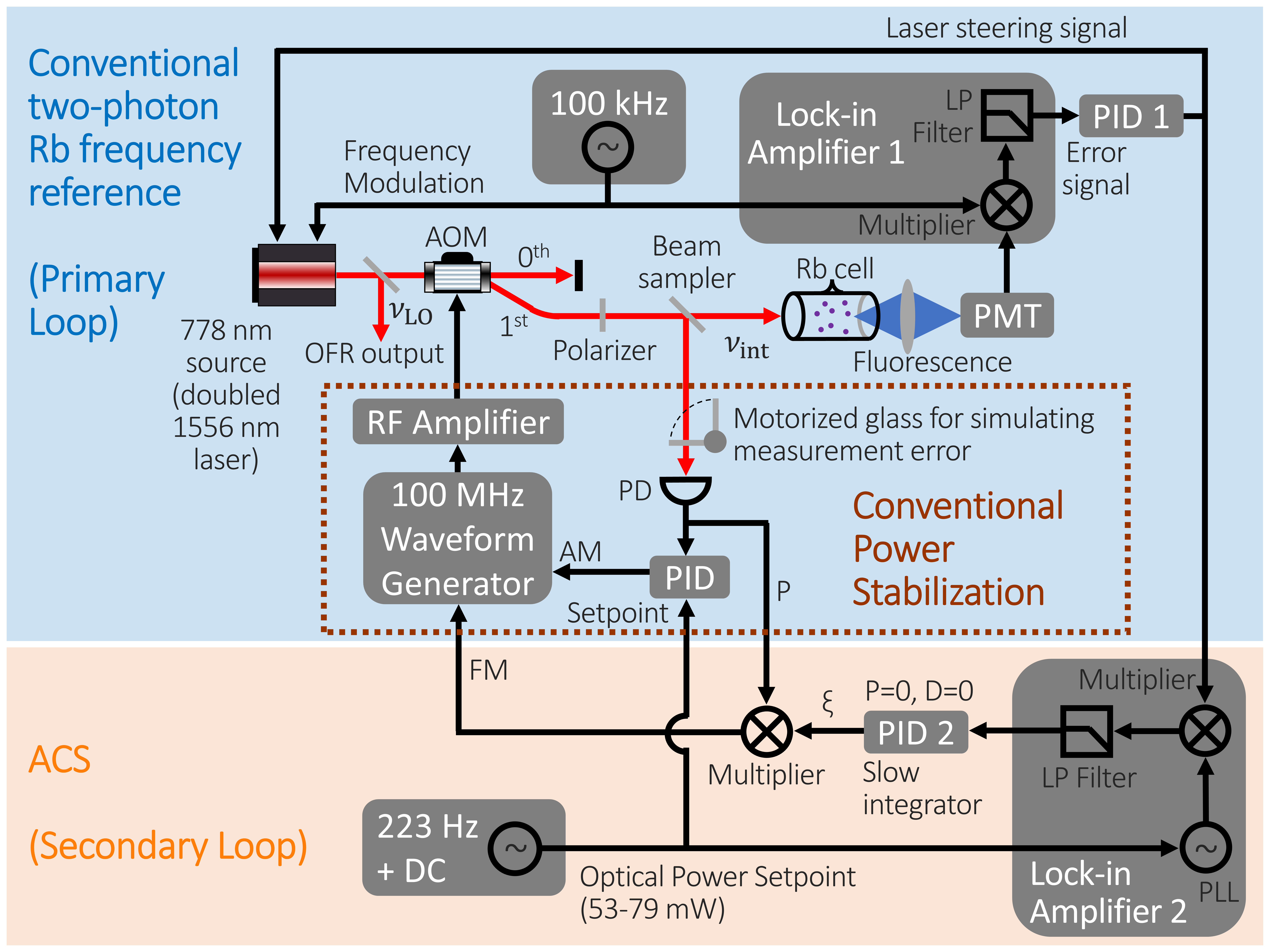}
    \caption{A block diagram representing our experimental implementation of ACS in a two-photon rubidium OFR. For simplicity, this diagram does not show the RAM suppression signal pathway or the optical frequency measurement setup used to characterize our OFR stability. The waveform generator and the multiplier connected to it are shown as separate instruments for clarity, but they are implemented in a single FPGA device. PD = photodiode; PMT = photomultiplier tube; PID = proportional/ integral/ derivative controller; AOM = acousto-optic modulator; PLL = phase locked loop; LP = lowpass.}
    \label{fig_exp_diagram}
\end{figure}

\section{Our OFR setup}
A block diagram of the experimental setup is shown in Fig. \ref{fig_exp_diagram}. Other than the components related to ACS (shown in the orange section), our architecture is typical for a degenerate-wavelength two-photon rubidium OFR. We use a narrow-linewidth frequency-doubled 1556 nm laser to generate light at 778.1 nm. This light is retroreflected to create two overlapping, counterpropagating excitation beams in a magnetically shielded, glass-blown 9-cm-long vapor cell. The cell contains isotopically enriched $^{87}$Rb and is maintained at 93 \degree C. When the light is resonant, the atoms are excited from the $5$S$_\frac{1}{2}$ state to the $5$D$_\frac{5}{2}$ state and emit fluorescence at 420 nm, which is separated from the excitation light by an interference filter, collected by a lens, and detected by a photomultiplier tube. We lock our laser to the $F=2 \rightarrow F=4$ transition with a bandwidth of about 2 kHz. The error signal for this lock is derived by modulating the laser frequency at $f_\text{FM} = 100$ kHz via direct laser current modulation and demodulating the fluorescence signal.

We used a camera to image the beam spot at several positions. The beam's $1/e^2$ radius varies from about 270 \unit{\micro\meter} to 310 \unit{\micro\meter} as the beam progresses through the vapor cell. We measure an AC Stark shift coefficient of about 720 Hz/mW, which is consistent with the expected value for our measured beam size \cite{martin_frequency_2019} when one accounts for the spatial distribution of optical power and the losses from reflections at the vapor cell windows.

We use a fiber-coupled acousto-optic modulator (AOM) both to control the optical power and to introduce the optical frequency shift required for ACS. The output fiber is coupled to the first-order diffracted beam, so by controlling the amplitude and frequency of the AOM's RF drive, both the power and the frequency shift on the output can be adjusted. The AOM's RF drive is generated by a field programmable gate array (FPGA)-based digital signal processor (DSP). In our experiment, the DSP is referenced to a hydrogen maser for frequency stability, but in a field-deployable atomic clock, it can also be referenced to the clock's own output.

To measure and stabilize the optical power entering the vapor cell, a linear polarizer and a wedged, fused silica beam sampler are placed just before the cell, diverting a small fraction of the light onto a photodiode. During conventional operation, the measured optical power $P$ is stabilized to 66 mW, and the AOM's RF drive frequency is held constant. During ACS operation, the power setpoint is varied sinusoidally at $f_\text{PM} = 223$ \unit{\hertz} around an average value $P_0$ of 66 mW. Our choice of $f_\text{PM} = 223$ \unit{\hertz} is motivated in the Supplemental Material \cite{sup_mat}. Except where specified otherwise, the power is modulated from 53 mW to 79 mW ($A = 0.19$). We observe that the clock shift due to the AC Stark effect has a linear relation with probe power over this range of operating parameters to within our measurement uncertainty. A digital lock-in amplifier monitors the resulting response on the laser steering signal. Its output is digitally integrated to yield $\xi$, which is passed to the AOM drive FPGA together with $P$ so that it can implement Eq. (\ref{eq_shift}). The fully digital implementation of our ACS signal pathway ensures that ACS is not compromised by voltage offsets between instruments. We did not take special care to measure or minimize the voltage offset on the power measurement photodiode. Such an offset would cause the ACS-enabled frequency reference to retain a small residual sensitivity to the AC Stark effect; however, we demonstrate a thousandfold suppression of the AC Stark shift sensitivity without addressing this offset.

Residual amplitude modulation (RAM) is reduced by active cancellation with a tone at $f_{FM} = 100$ kHz applied to the AOM RF drive power. The phase and amplitude of the tone are controlled by feedback loops in order to zero the RAM as measured on the power measurement photodiode, a technique known as complex modulation \cite{aronson_reduction_2020, chia_complex_2022, nguyen_field-programmable_2024}. Out-of-loop measurements confirm that our clock stability is not limited by RAM fluctuations.

To characterize the OFR's stability, we pick off some light from the stabilized 1556 nm laser and heterodyne it with an optical frequency comb stabilized to an ultrastable optical reference cavity. The beat frequency is measured by a counter. To account for long-term drift of the cavity (about $10^{-13}/\text{h}$), the comb's repetition rate is measured by a phase noise analyzer referenced to a hydrogen maser, and the drift is subtracted off in postprocessing.

\begin{figure}
    \centering
    \includegraphics[width=3.4in]{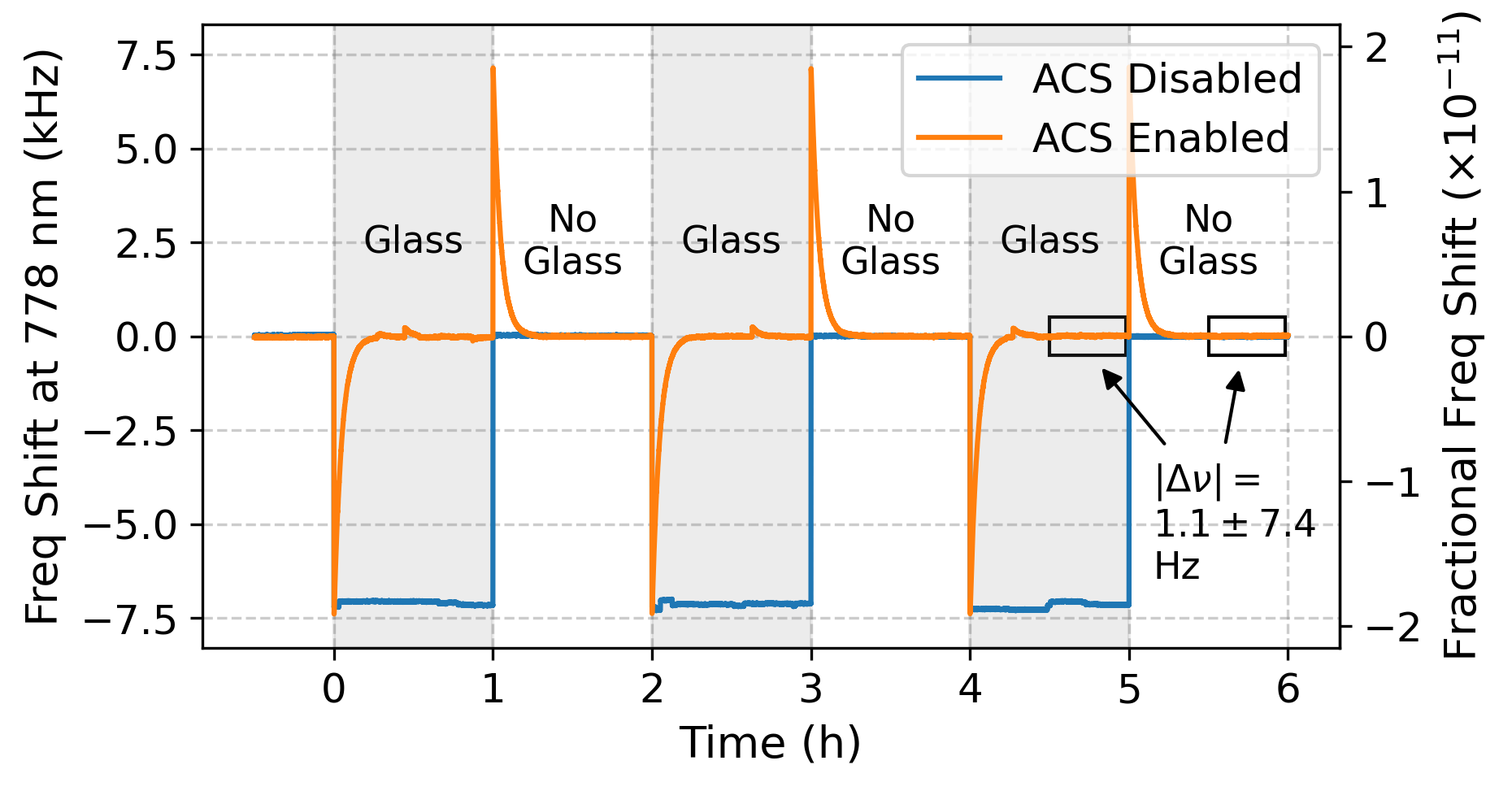}
    \caption{The response of our OFR to intentional optical power variations, caused by alternatingly raising and lowering a piece of glass in front of the power measurement photodiode once per hour. ACS suppresses the OFR response by a factor of about 1000; the residual shift on each step is near or below the 7.4 Hz uncertainty (Allan deviation at $\tau = 30 \text{ min}$). For the ACS data, $T = 170$ s.}
    \label{fig_step_test}
\end{figure}

\section{Characterization of ACS performance}
We performed two experiments to characterize the performance of the ACS method. First, we demonstrated that ACS dramatically reduces the OFR's sensitivity to optical power variations by intentionally applying large, discrete steps in the power. To apply the power variations, we used a servo motor to alternatingly raise and lower a piece of glass once per hour in front of the photodiode responsible for stabilizing the optical power, as shown in Fig. \ref{fig_exp_diagram}. This simulates a multiplicative error on the power measurement, similar to one that might naturally arise due to variations in the gain of the photodiode, the reflectance of the beam sampler, or the transmittance of the vapor cell window. We performed five cycles of inserting and removing the glass, i.e. ten total power steps. The results from three cycles are shown in Fig. \ref{fig_step_test}.

During regular operation (ACS disabled), the insertion or removal of the glass causes a frequency step of about 7 kHz at 778 nm ($1.8\times10^{-11}$). When ACS is enabled, the initial response is the same, but the slow secondary feedback loop then adjusts $\xi$ until the OFR returns to very nearly its original frequency. The response time $T$ of the secondary feedback loop, defined as the $1/e$ decay time of $\xi$ after an impulse, is set to about 170 s for this experiment. Note that during the intervals in which glass is in place in front of the photodiode, our data show small frequency fluctuations regardless of whether ACS is enabled. This is because the glass was taped, rather than rigidly mounted, to the motor arm, and therefore occasionally suffered small movements. ACS even corrects these small, unintentional power fluctuations; they are visibly transient with ACS enabled and persistent with ACS disabled.

We calculated residual shifts for each ACS-enabled power step using data starting 30 minutes (about 10 response times) after the step, excluding any data that visually appears to be affected by the transient glass-induced fluctuations. The residual shift can be measured to a resolution of approximately 7.4 Hz, as this is the OFR's Allan deviation at $\tau = 30$ min. Six of the ten measured power steps exhibited residual shifts below $\pm 7.4$ Hz, i.e. consistent with 0, indicating a thousandfold suppression of the OFR response compared to when ACS is disabled. The remaining four had residual shifts of up to $\pm 20$ Hz, which we attribute to difficulty in isolating data free from glass fluctuations.

\begin{figure}
    \centering
    \includegraphics[width=3.4in]{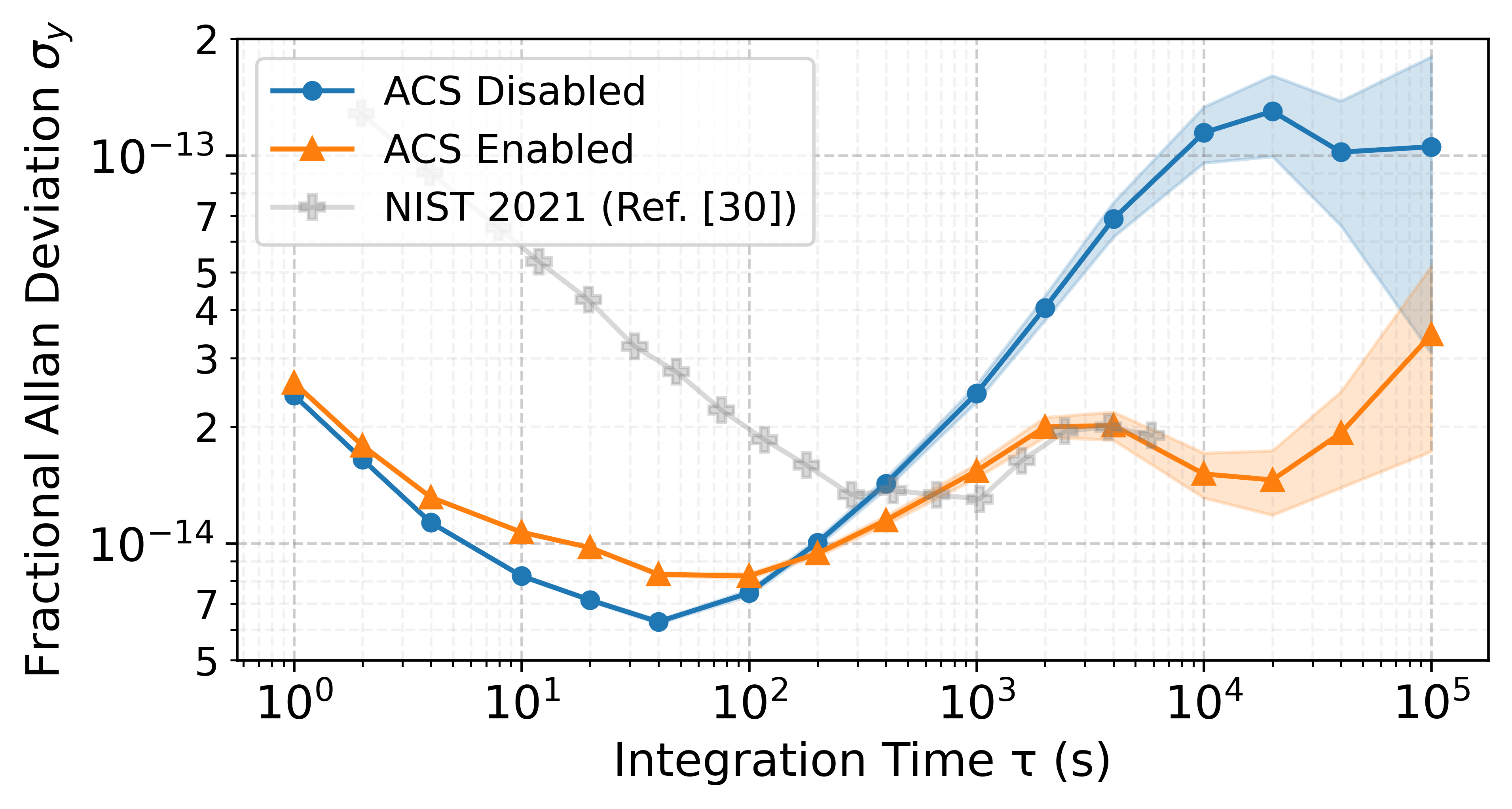}
    \caption{Allan deviations of the OFR with and without ACS. Shaded regions denote standard error. For the ACS data, $A = 0.19$ and $T \approx 200 \text{ s}$. Our previously reported two-photon rubidium performance, Ref. \cite{newman_high-performance_2021}, is shown in gray for context. (Note that for the data shown here, a nanoparticle polarizer was used before the power monitor pickoff, while in Figs. \ref{fig_step_test} and \ref{fig_adev_transitions}, a polarizing beamsplitter cube was used. We believe this to be responsible for the slight difference in the shapes of the Allan deviations.)}
    \label{fig_adevs}
\end{figure}

Next, we investigated the frequency stability that can be obtained with ACS. We operated the OFR for several days each with ACS and with conventional power stabilization. No intentional power variations were applied during this experiment. The frequency stability observed during these runs is shown in Fig. \ref{fig_adevs}. Our intense excitation laser enables a short-term fractional Allan deviation below $3 \times 10^{-14}$ at 1 s, seven times better than the short term stability previously reported in our two-photon rubidium OFR in Ref. \cite{newman_high-performance_2021}. When the OFR is operated conventionally, our use of intense light comes at the expense of stability at longer integration times, but when ACS is enabled, the Allan deviation past $10^2$ s is improved to a level comparable to that reported in Ref. \cite{newman_high-performance_2021}, remaining at or below $2 \times 10^{-14}$ until $4 \times 10^4$ s. It is currently unclear what limits the long-term stability of our ACS-enabled frequency reference. One possible source of instability is fluctuations on the photodiode voltage offset. This offset causes a power measurement error that is additive rather than multiplicative, so it causes an AC Stark shift that ACS is unable to compensate.

The exact mechanism for the larger AC Stark shift fluctuations that limit the performance during conventional operation is also currently unclear. This instability appears to be bounded in nature. We suspect an environmental sensitivity in the polarizer just before the power measurement pickoff; the data in Fig. \ref{fig_adevs} was collected using a nanoparticle polarizer (colorPol VIS 700 BC4 CW02 with antireflection coating), but the stability at integration times between $10^2$ s and $10^3$ s was improved somewhat by replacing this polarizer with a polarizing beamsplitter cube (New Focus 5812). Allan deviations collected with the polarizing beamsplitter cube are shown in Fig. \ref{fig_adev_transitions}. However, out of all our measurements, the ones using ACS in conjunction with the nanoparticle polarizer consistently yielded the lowest Allan deviations beyond $\tau = 10^4$ s. Multiple measurements alternating between the two polarizers indicate that these observations are repeatable. Both polarizers were angled for slightly off-normal incidence to reduce interference from stray reflections.

\section{Stability limit imposed by LO noise}
In this section, we examine how ACS provides a new avenue for frequency noise on the LO to couple onto the clock's stable output. As shown in Fig. \ref{fig_abstracted_diagram}, the ACS method relies on the detection of a tone at $f_\text{PM}$ on the LO steering signal, i.e. the output of the primary feedback loop. It is therefore inherently sensitive to any noise on the steering signal containing a spectral component at $f_\text{PM}$. There are two contributions to such noise: frequency noise $S_{\nu, \text{free}}$ on the free-running LO, which must be corrected by the steering signal in order to maintain lock, and noise $S_{\nu, \text{servo}}$ introduced incidentally by the primary feedback loop, such as photon shot noise or technical noise, which typically limits the locked LO's short-term stability. In general, $S_{\nu, \text{free}}$ is likely to dominate over $S_{\nu, \text{servo}}$, but we use a quiet LO and must therefore consider both contributions. Noise at $f_\text{PM}$ from either of these sources is detected by the secondary feedback loop and results in noise on $\xi$ at timescales greater than $T$, the response time of the secondary feedback loop. This in turn causes noise on the locked clock frequency $\nu_\text{LO}$, imposing a limit on the clock performance at integration times $\tau > T$. Around $\tau = T$, the fractional Allan deviation $\sigma_y$ must transition to a second, degraded line given by
\begin{equation}
\label{eq_stability_limit}
    \sigma_y(\tau > T) = \frac{1}{\nu_\text{LO}} \sqrt{\frac{S_{\nu, \text{free}}(f_\text{PM}) + S_{\nu, \text{servo}}(f_\text{PM})}{\tau}} \frac{1}{A},
\end{equation}
where $A$ is the dimensionless amplitude of the power modulation defined by Eq. (\ref{eq_power_modulation}) and $S_{\nu, \text{free}}$ and $S_{\nu, \text{servo}}$ are one-sided power spectral densities with units of LO frequency squared per unit of spectral bandwidth. Eq. (\ref{eq_stability_limit}) is derived in the Supplemental Material \cite{sup_mat} (see also Ref. \cite{rutman_characterization_1991} cited therein).

\begin{figure}
    \centering
    \includegraphics[width=3.4in]{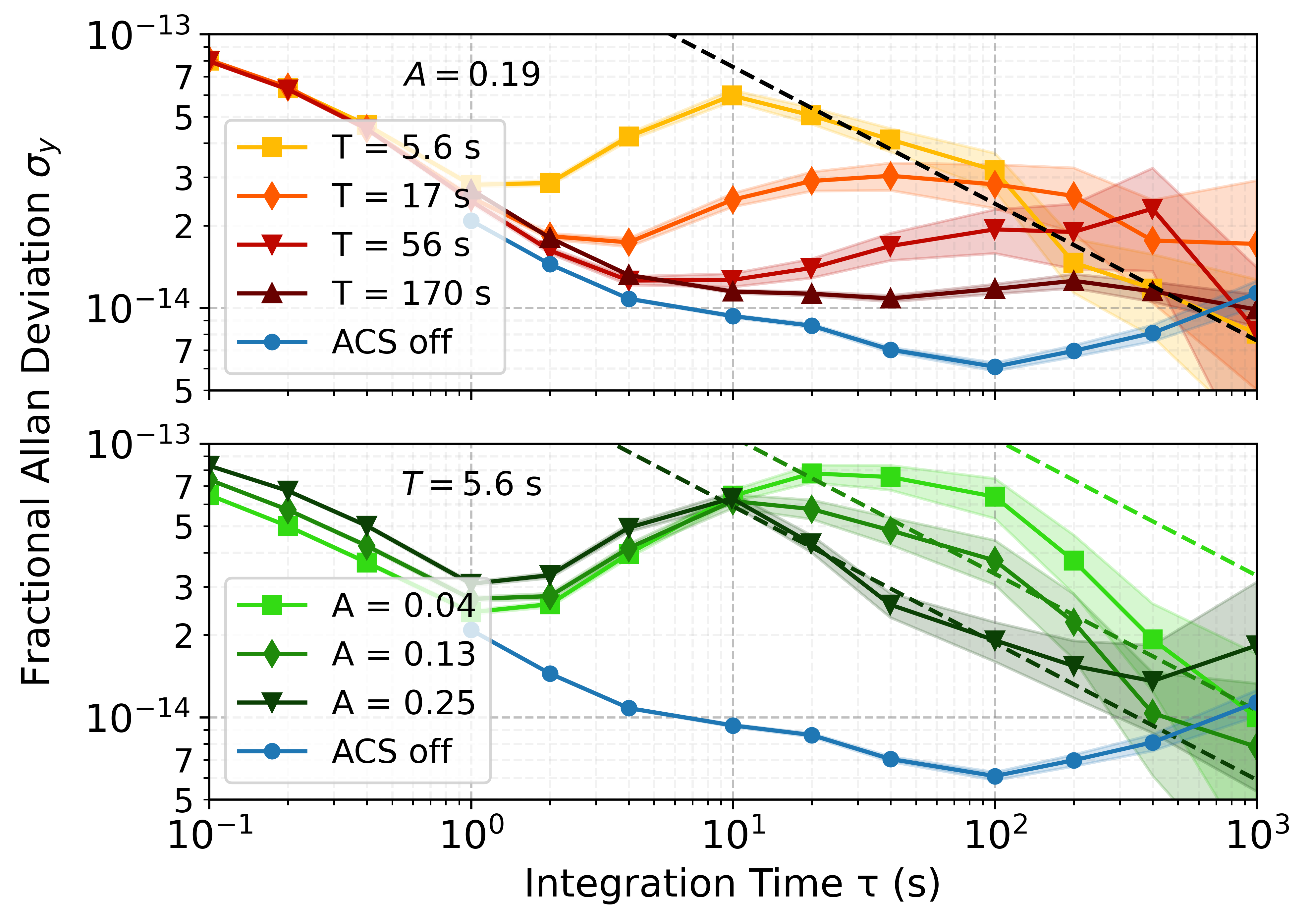}
    \caption{Allan deviations of our OFR for different choices of the ACS parameters $A$ and $T$. Shaded regions denote standard error. For integration times $\tau \gg T$, the Allan deviations show reasonable agreement with the stability limits predicted by Eq. (\ref{eq_stability_limit}), shown in dashes. For these measurements, we chose time constants considerably faster than the optimal value of about 200 s to ensure that the OFR stability is governed by this limit, without contributions from systematic instabilities that ultimately limit performance at the longest integration times.}
    \label{fig_adev_transitions}
\end{figure}

To investigate this stability limit imposed on ACS, we performed several additional measurements with different choices for the parameters $A$ and $T$. The Allan deviations obtained in these runs are shown in Fig. \ref{fig_adev_transitions}. Dashed lines show the theoretical Allan deviations predicted by Eq. (\ref{eq_stability_limit}). To determine $S_{\nu, \text{free}}(f_\text{PM})$ for the calculation of the dashed lines, we measured the frequency noise spectrum of the free-running laser (plotted in the Supplemental Material \cite{sup_mat}). $S_{\nu, \text{servo}}$ is assumed to be white noise and is estimated from the measured short-term stability of the locked OFR (specifically, the Allan deviation at $\tau = 0.1 \text{ s}$) for each curve. Inaccuracies in estimating these noise terms are likely responsible for the small disagreements between theory and experiment in Fig. \ref{fig_adev_transitions}.

Our stability limit is analogous to the intermodulation effect \cite{audoin_limit_1991}, differing only in that it occurs on the secondary feedback loop instead of the primary one; it is caused by aliasing between the power modulation and noise on the LO. Almost all two-loop methods rely in some form on periodic modulation of the interrogation in order to generate a secondary error signal, and are therefore subject to similar limits, though the limit might be very low if the local oscillator is stable on the timescale of the modulation.

The stability limit given by Eq. (\ref{eq_stability_limit}) can be optimized by using a low-noise LO as well as appropriate choices of the parameters $f_\text{PM}$, $A$, and $T$. The modulation frequency $f_\text{PM}$ should be chosen in a quiet part of the LO's noise spectrum, which is typically at higher frequencies, though it must be much slower than the bandwidth of the primary frequency stabilization feedback loop. The modulation amplitude $A$ should be as large as possible while ensuring that the measured AC Stark shift remains linear over the range of powers employed, though it is also worth considering that the short-term performance will be slightly degraded for large values of $A$ due to the time spent at lower optical power, as shown in Fig. \ref{fig_adev_transitions} (lower). The response time $T$ of the secondary feedback loop should be chosen to be slightly faster the timescale at which the AC Stark shift begins to limit the OFR's stability, which is typically a few hours or days; if $T$ is slower than this timescale, ACS fails to keep up with AC Stark shift fluctuations, and if it is much faster, the transition to the degraded Allan deviation happens unnecessarily quickly.

In order to enable ACS to improve the frequency stability of our OFR, it was essential to choose a low-noise laser so that the degraded stability limit lies below the AC Stark shift limit. For this reason, we chose a commercially available whispering gallery mode self-injection-locked laser at 1556 nm. Its noise spectrum is shown in the Supplemental Material \cite{sup_mat}. The ongoing development of compact and integrated low-noise optical lasers \cite{nader_heterogeneous_2025, isichenko_sub-hz_2024, lu_emerging_2024, di_gaetano_7781_2024, zhang_photonic_2023, tran_extending_2022} suggest that it may be possible for field-deployable OFR's to benefit from ACS or similar two-loop methods.

An approach has been described to suppress the contribution of oscillator noise, $S_{\nu, \text{free}}(f_\text{PM})$, to the stability limit described by Eq. (\ref{eq_stability_limit}) \cite{gerginov_two-photon_2018}. It entails a differential measurement from two simultaneous atomic interrogation regions, one with applied power modulation and one without. In future work, it would be interesting to explore whether this approach can enable additional stability improvements.

\section{Discussion}
We have demonstrated that ACS reduces the AC Stark shift sensitivity of our two-photon rubidium OFR, alleviating the tradeoff between short-term stability and long-term stability. In its current form, ACS requires only an AOM and digital signal processing. In order to reduce power consumption, future work could explore alternative two-loop methods that do not require an AOM. Such alternative methods may be useful for field-deployable devices, which are subject to environmental perturbations that make it difficult to stabilize operating parameters such as interrogation power at long timescales. Our implementation of ACS in a two-photon rubidium OFR demonstrates that two-loop methods have the potential to improve the stability of cw optical clocks, which are poised to become widely adopted for applications in navigation, communication, and sensing in the coming decade.

\section*{Acknowledgments}
This research is partially supported by the National Institute of Standards and Technology (NIST) and the Office of Naval Research (ONR). Y.A. acknowledges support under the Professional Research Experience Program (PREP), funded by the National Institute of Standards and Technology and administered through the Department of Physics, University of Colorado, Boulder. We thank Gregory Hoth and Zachary Newman for providing valuable feedback on the manuscript. Any mention of commercial products within this letter is for information only; it does not imply recommendation or endorsement by NIST.

\bibliography{references}

\end{document}


\title{\textbf{Supplemental Material - Active compensation of the AC Stark shift in a two-photon rubidium optical frequency reference using power modulation}
}

\author{Yorick Andeweg}
\affiliation{National Institute of Standards and Technology, Boulder, Colorado, United States}
\affiliation{University of Colorado Boulder, Boulder, Colorado, United States}

\author{John Kitching}
\affiliation{National Institute of Standards and Technology, Boulder, Colorado, United States}
\author{Matthew T. Hummon}
\affiliation{National Institute of Standards and Technology, Boulder, Colorado, United States}

\maketitle

\begin{figure}[H]
    \centering
    \includegraphics[width=3.4in]{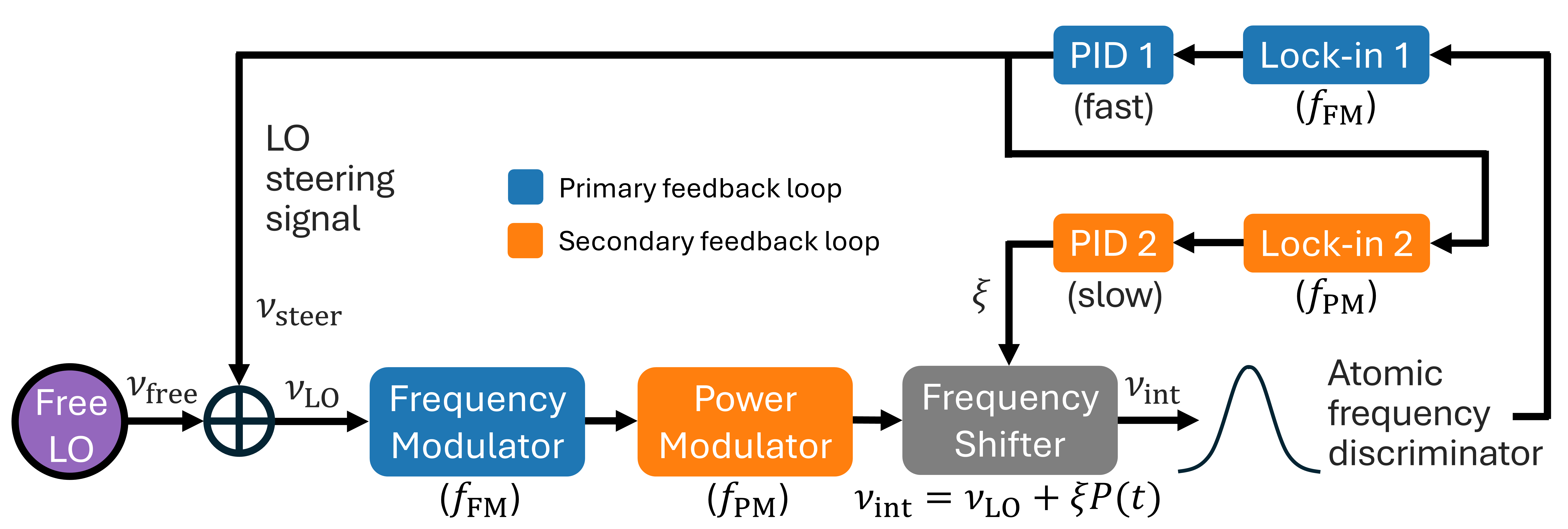}
    \caption{A high-level block diagram of the ACS method for suppressing the AC Stark shift in a generic CW frequency reference. This diagram is identical to Fig. 1 in the main text, except that here we explicitly show the locked oscillator frequency $\nu_\text{LO}$ as the sum of the LO free-running frequency $\nu_\text{free}$ and the LO steering signal $\nu_\text{steer}$ applied to maintain lock, as our stability limit derivation refers to these frequencies.}
    \label{fig_supmat_abstracted_diagram}
\end{figure}
\textit{Derivation of ACS stability limit}---This section presents a derivation of the stability limit imposed on clocks employing the auto-compensated shift (ACS), Eq. (3) in the main text. Fig. \ref{fig_supmat_abstracted_diagram} shows a block diagram of the ACS method, with labels indicating the various frequencies that appear in the derivation. This derivation has the following parts:
\begin{enumerate}[(a)]
\item Derive an expression for the LO steering signal $\nu_\text{steer}$ that stabilizes the locked oscillator frequency $\nu_\text{LO}$, including both the effect of the applied modulation and two noisy terms, $\nu_\text{free}$ and $\nu_\text{servo noise}$, which have spectral components at $f_\text{PM}$.
\item Describe how the two noisy terms lead to low-frequency noise on $\xi$.
\item Describe how this noise on $\xi$ translates to noise on the OFR's stable output frequency, $\nu_\text{LO}$.
\item Assign power spectral densities $S_{\nu, \text{free}}$ and $S_{\nu, \text{servo}}$ to the noisy terms in part (a) and propagate these fluctuations to the noise on $\nu_\text{LO}$.
\end{enumerate}

\noindent (a) In ACS, the optical power $P$ is modulated at a frequency $f_\text{PM}$:
\begin{equation}
\label{eq_power_modulation_sup}
    P(t) = P_0 [1 + A \sin(2 \pi f_\text{PM} t)].
\end{equation}
This perturbs the atomic transition frequency $\nu_\text{trans}$:
\begin{equation}
\label{eq_light_shift}
    \nu_\text{trans}(t) = \nu_0 + c P(t)
\end{equation}
where $\nu_0$ is the unperturbed transition frequency and $c$ is the AC Stark shift coefficient. Let us follow the modulation effect upstream to $\nu_\text{steer}$. The primary feedback loop locks the interrogation frequency $\nu_\text{int}$ to the transition frequency with some noise:
\begin{equation}
\label{eq_servo_lock}
    \nu_\text{int}(t) = \nu_\text{trans}(t) - \nu_\text{servo noise}(t)
\end{equation}
where $\nu_\text{servo noise}(t)$ is a rapidly-fluctuating term that averages to zero, representing photon shot noise, technical noise, or any other noise source that limits the short-term performance of the primary feedback loop. (The sign on $\nu_\text{servo noise}(t)$ is arbitrary; we make it negative here to aid the clarity later on in the derivation.) Substituting Eq. (\ref{eq_servo_lock}) into Eq. (\ref{eq_light_shift}) and using the fact that $\nu_\text{int} = \nu_\text{LO} + \xi P$, we find
\begin{equation}
\label{eq_vint_lock}
\begin{split}
    \nu_\text{LO} &= \nu_0 + c P(t) - \xi P(t) - \nu_\text{servo noise}(t)
\end{split}
\end{equation}
Next, we can  write the locked oscillator frequency $\nu_\text{LO}$ as the sum of the rapidly fluctuating free-running LO frequency $\nu_\text{free}(t)$ and a steering term $\nu_\text{steer}(t)$ provided by the fast primary feedback loop. Substituting this into Eq. (\ref{eq_vint_lock}) and solving for the steering term yields
\begin{equation}
\label{eq_nu_steer}
    \nu_\text{steer}(t) = \nu_0 + c P(t) - \xi P(t) - \nu_\text{servo noise}(t) - \nu_\text{free}(t).
\end{equation}

\noindent (b) The secondary feedback loop uses a lock-in amplifier to detect a tone at $f_\text{PM}$ on $\nu_\text{steer}(t)$, given by Eq. (\ref{eq_nu_steer}). Clearly, this includes the terms containing $P(t)$, but perhaps less obviously, it can include the rapidly fluctuating terms $\nu_\text{free}$ and $\nu_\text{servo noise}$ as well. Suppose that these terms each have a spectral component at $f_\text{PM}$:
\begin{subequations}
\label{eq_define_pqrs}
\begin{align}
    \nu_\text{free}(t) = &p \sin(2 \pi f_\text{PM} t) + q \cos(2 \pi f_\text{PM} t)\notag\\
    &+ \text{other spectral components} \label{eq_define_pq}\\
    \nu_\text{servo noise}(t) = &r \sin(2 \pi f_\text{PM} t) + s \cos(2 \pi f_\text{PM} t)\notag\\
    &+ \text{other spectral components} \label{eq_define_rs}
\end{align}
\end{subequations}
where the signed amplitudes $p$, $q$, $r$, and $s$ vary slowly compared to $f_\text{PM}$ and average to zero. The lock-in amplifier in the secondary feedback loop does not detect the cosine terms, since they are 90\degree out of phase with the applied perturbation (Eq. (\ref{eq_power_modulation_sup})), but it detects the sine terms. The complete signal detected by this lock-in amplifier is as follows:
\begin{equation}
    \nu_\text{steer, detected} = (c P_0 A - \xi P_0 A - r - p) \sin(2 \pi f_\text{PM} t)
\end{equation}
The secondary feedback loop tunes $\xi$ to eliminate this detected tone, yielding
\begin{equation}
\label{eq_xi}
    \xi = c - \frac{r}{P_0 A} - \frac{p}{P_0 A}.
\end{equation}
Recall that the secondary feedback loop is intended to tune $\xi$ to match $c$ in order to render $\nu_\text{LO}$ insensitive to power fluctuations. The additional terms represent slowly-fluctuating noise.

\noindent (c) Let us see how this noise on $\xi$ perturbs the stable output frequency, $\nu_\text{LO}$. Substituting Eq. (\ref{eq_xi}) into Eq. (\ref{eq_vint_lock}) yields
\begin{equation}
\label{eq_tone_and_error}
\begin{split}
    \nu_\text{LO} &=
    \begin{aligned}[t]
    \nu_0 &+ c P(t) - \left(c - \frac{r}{P_0 A} - \frac{p}{P_0 A}\right) P(t)\\
    &- \nu_\text{servo noise}(t)
    \end{aligned}\\
    &=
    \begin{aligned}[t]
    \nu_0 &+ \frac{p}{A} + p \sin(2 \pi f_\text{PM} t)\\
    &+ \frac{r}{A} + r \sin(2 \pi f_\text{PM} t) - \nu_\text{servo noise}(t)
    \end{aligned}\\
\end{split}
\end{equation}
using Eq. (\ref{eq_power_modulation_sup}) for the expression $P(t)$. The terms $\frac{p}{A}$ and $\frac{r}{A}$ cause the greatest perturbations, since $A < 1$. Let us see how these terms fluctuate at long times.

\noindent (d) Suppose that $\nu_\text{free}$ and $\nu_\text{servo noise}$ have one-sided power spectral densities (PSD's) $S_{\nu, \text{free}}$ and $S_{\nu, \text{servo}}$ respectively. The amplitudes $p$ and $r$ can then be said to have PSD's of their own at frequencies below the bandwidth $1/T$ of the secondary feedback loop, given by
\begin{equation}
\begin{split}
    S_p(f < 1/T) &= 2 S_{\nu, \text{free}} (f_\text{PM})\\
    S_r(f < 1/T) &= 2 S_{\nu, \text{servo}} (f_\text{PM}),
\end{split}
\end{equation}
as described in Ref. \cite{audoin_limit_1991}. The PSD of the troublesome frequency terms $\frac{p}{A} + \frac{r}{A}$ on $\nu_\text{LO}$ is then
\begin{equation}
\label{eq_error_noise}
\begin{split}
    &S_{\nu, \text{LO}}(f<1/T)\\
    &= \frac{1}{A^2} [S_p(f<1/T) + S_r(f<1/T)]\\
    &= \frac{2}{A^2}[S_{\nu, \text{free}}(f_\text{PM}) + S_{\nu, \text{servo}}(f_\text{PM})]
\end{split}
\end{equation}
Substituting this into the standard formula for the Allan deviation of a white noise spectrum \cite{rutman_characterization_1991}, and dividing by $\nu_\text{LO}$ to make it fractional, we arrive at the stability limit equation we present in the main text:
\begin{equation}
\label{eq_stability_limit_sup}
    \sigma_y = \frac{1}{\nu_\text{LO}} \sqrt{\frac{S_{\nu, \text{free}}(f_\text{PM}) + S_{\nu, \text{servo}}(f_\text{PM})}{\tau}} \frac{1}{A}.
\end{equation}

\begin{figure}
    \centering
    \includegraphics[width=3.4in]{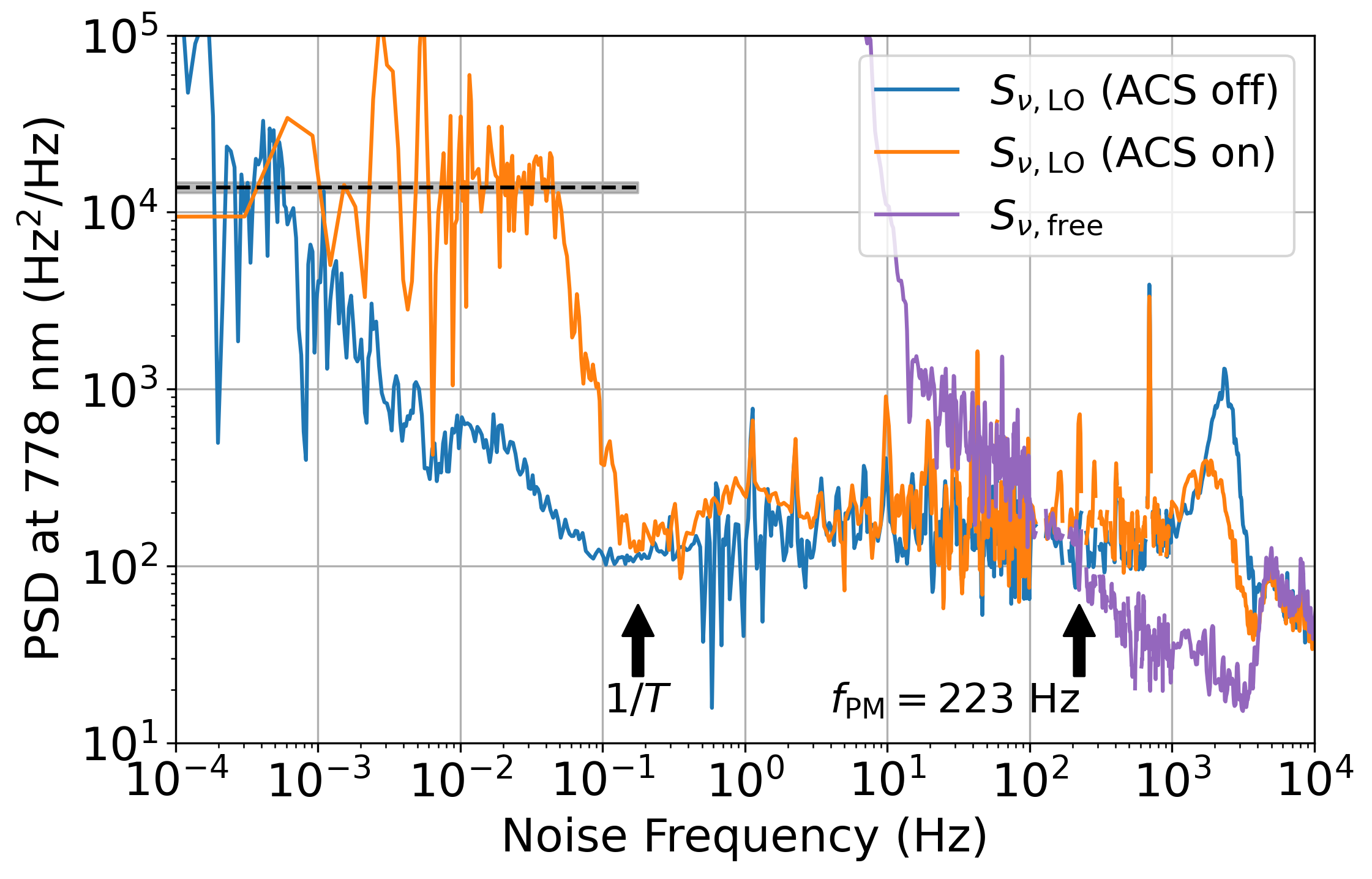}
    \caption{The frequency noise spectrum (one-sided PSD) of our laser. The horizontal dashed line shows the white noise level predicted at low frequencies by Eq. (\ref{eq_error_noise}). For the ACS data shown here, $T = 5.6 \text{ s}$ and $A = 0.19$.}
    \label{fig_noise_spectra}
\end{figure}

\textit{Frequency noise spectrum of our laser}---Fig. \ref{fig_noise_spectra} shows the frequency noise spectrum of our laser when it is unlocked, locked with ACS, and locked without ACS. For noise frequencies below 10 Hz, we obtain this spectrum by heterodyning our clock laser with a frequency comb, measuring the beat frequency on a counter, and performing a Fourier transform during postprocessing. To measure the noise spectrum at frequencies greater than 10 Hz, we pass the beat note through a frequency-to-voltage converter and into an FFT analyzer. The small gaps in the data are where we removed spikes at harmonics of the power line frequency (60 Hz). 

The purple curve represents $S_{\nu, \text{free}}$, the noise spectrum of the free-running laser. The blue curve shows the noise spectrum of the locked laser $S_{\nu, \text{LO}}$ with only the fast primary feedback loop active (ACS off).  From comparison of these two spectra we can observe the servo bump of the fast primary feedback loop around $2\times10^3$ Hz, and a white-noise floor of the primary servo near $10^2~\text{Hz}^2/\text{Hz}$.  To estimate the long-term stability limit for the laser when ACS is enabled using Eqs. (11) and (12), we can estimate the value $S_{\nu, \text{free}}(f_\text{PM})$ directly from the free running laser PSD, and $S_{\nu, \text{servo}}(f_\text{PM}) =S_{\nu, \text{LO}}(f_\text{PM})$ when ACS is off.  At $f_\text{PM} = 223 \text{ Hz}$, these curves have values of $125 \pm 8 \text{ Hz}^2/\text{Hz}$ and $130 \pm 9 \text{ Hz}^2/\text{Hz}$ respectively, determined by averaging data points in a $\pm 30 \text{ Hz}$ window around $223 \text{ Hz}$. Substituting these values into Eq. (\ref{eq_error_noise}) gives the predicted white noise level for ACS at low frequencies ($f < 1/T$), shown by the dashed line in Fig. \ref{fig_noise_spectra}. The uncertainty intervals in this analysis, shown graphically by a thin shaded region around the dashed line, denote standard error.

Fig. \ref{fig_noise_spectra} motivates our choice of $223 \text{ Hz}$ for $f_\text{PM}$. Choosing a lower frequency would worsen the ACS stability limit because of the higher value of $S_{\nu, \text{free}}$, but choosing a higher frequency would offer little improvement, as the stability limit would become limited by $S_{\nu, \text{servo}}$. We chose a prime number for $f_\text{PM}$ to avoid aliasing with other frequencies in the lab.

\bibliography{references}